# An easy way to Gravimagnetism

By Prof. Dr. Claus W. Turtur
University of Applied Sciences Braunschweig- Wolfenbüttel
Salzdahlumer Straße 46 / 48
GERMANY - 38302 Wolfenbüttel
Phone: (++49) 5331 / 939 - 3412
E-mail.: c-w.turtur@fh-wolfenbuettel.de

**Abstract:**
One of the numerous results of the Theory of General Relativity is the appearance of Gravimagnetism, described already by Thirring and Lense in 1918 [THI 18]. Its understanding normally remains reserved to specialists.
This article presents a simple classical analogy to bring the description of Gravimagnetism within the reach of beginners. The article describes a classical model calculation on the level of undergraduate students and mentions the situation of the modern experiment.

**Structure of the article:**
(1.)   Qualitative description of the basics
(2.)   An analogy between gravitation and electricity
(3.)   The example of a gyrator in the gravimagnetic field of the earth
(4.)   References of literature

## Part 1.: Qualitative description of the basics

Regarding electromagnetic interaction, it is well known, that it contains a static an a dynamic part. The electrostatic component is described by electrical charge according to Coulomb's law. The electrodynamic (=electromagnetic) component is described by the use of moving charge following Biot- Savart's law. [ABE 71/81]. We also know the Lorentz- force and Ampere's law.

The difference between electrostatic and electromagnetic forces is merely a transformation of coordinates. This means, if an electrical test- charge is moving exactly synchronously with an other field- producing charge, the test- charge will only notice an electrostatic field but no magnetic one. If the test- charge will move relatively to the field- producing charge, it will remark additionally an electromagnetic field and force.

Because the transformation of coordinates is of central importance in electrodynamics same as in gravitation, the idea might arise, that the ambivalence of static and dynamic forces can be transferred from electricity to gravitation. As we will see in the following pages, this is really the case. The analogy is illustrated in table 1. On the following pages we will see, that this analogy is reaching far enough to describe gravimagnetism.

| | electric forces | gravitational forces |
|---|---|---|
| **static forces** | electrostatic interaction (described by Coulomb) | gravitostatic interaction (described by Newton) |
| **dynamic forces (magnetic)** | electromagnetic interaction (described by Biot- Savart and Lorentz) | gravimagnetic interaction (described by Einstein, Thirring- Lense- Effekt) |
| Table 1.: Illustration of the analogy between static and dynamic forces regarding electricity and gravitation | | |

The calculations according to Coulomb, Biot- Savart and Newton are generally known and experimentally verified so perfect, that it is even standard for every technician. With Gravimagnetism, known as Thirring- Lense- Effect, knowledge behaves totally different. This effect is known mainly by specialsts and is normally calculated with the means of the theory of General Relativity (see for instance [GOE 96] or [SCH 02]). A further problem is also, that this effect can hardly be detected, because its forces are extremely small.

Indeed the experimental verification of Gravimagnetism is not acchieved until today. Worldwide there is nowadays only one experiment with a serious hope to give results within foreseeable future. This experiment is known under the name of "Gravity- Probe- B Experiment" [GPB 04]. In order to suppress the influence of gravitostatic forces, it has to be performed in a satellite, and it uses gyroscopes with a measuring accuracy 8 orders of magnitude (!) higher than the best gyroscopes in navigation satellites (see [DIT 99]).

An other experiment by the European Space- Agency (ESA) [HYP 00] is only in the phase of planning up to now. On the long run, it promises higher accuracy than the mechanical Gravity- Probe- B Experiment, because it is based on atomic beam interferometry. But it will still need several years of preparation.

Having the immense efforts and expenses in mind, which science devotes to the verification of Gravimagnetism, we understand two aspects: On the one hand the effect is of very fundamental importance for the basic understanding of theory, on the other hand we see, that the experimental verification is very difficult.

In order to bring a generally comprehensible explanation of the Thirring- Lense- Effect, we can follow table 1 and transfer the formulas of electromagnetism into analogous formulas of gravitomagnetism. This is what we want to do now.

## Part 2.: An analogy between Gravitation and Electricity

### (2.a.) The analogy of static forces
This analogy is known since long time and recited in equation 1:

Electrostatic force $F_{es}$

$$\vec{F}_{es} = \frac{1}{4\pi\varepsilon_o} \frac{q_1 q_2}{|\vec{r}_1 - \vec{r}_2|^2} \cdot \vec{e}_{\Delta r}$$

(equation 1)

Gravitostatic force $F_{gs}$

$$\vec{F}_{gs} = -\gamma \cdot \frac{m_1 m_2}{|\vec{r}_1 - \vec{r}_2|^2} \cdot \vec{e}_{\Delta r}$$

with $q_1, q_2$ = electrical charges and $m_1, m_2$ = ponderable masses
they are the medium for each interaction
and $\vec{r}_1, \vec{r}_2$ = positions of the electrical charges respectively the ponderable masses
$\vec{e}_{\Delta r}$ = unit vector connecting the partners of interaction

as well as the factors of proportionality:

$$\frac{1}{4\pi\varepsilon_0} = 8.893 \cdot 10^9 \frac{Nm^2}{C^2} \qquad \text{respectively} \qquad \gamma = 6.67 \cdot 10^{-11} \frac{Nm^2}{kg^2}$$

"Negative" sign represents attractive forces, "positive" sign stands for repulsive forces.

Remark about the term of field:
A punctiform charge $q_1$ produces (if not moving) an electrostatic field

$$\vec{E}_1 = \frac{1}{4\pi\varepsilon_0} \cdot \frac{q_1}{|\vec{r}_1 - \vec{r}_2|^2} \cdot \vec{e}_{\Delta r}, \qquad \text{(equation 2)}$$

in which an other punctiform charge $q_2$ will experience the force $\vec{F}_2$ with $\vec{F}_2 = q_2 \cdot \vec{E}_1$.

Except for the factors of proportionality (and the fact that the names "mass" and "charge" sound different), the formulas of electrostatics and gravitostatics look absolutely identical. So we decide to transfer this identity also to the dynamic formulas, as following.

## (2.b) The analogy of the magnetic (dynamic) forces

If an electric charge $q_2$ is moving relatively to a charge $q_1$, additionally to the static field $\vec{E}$ it will recognize a magnetic field $\vec{H}$. The latter one can be calculated according to figure 1 and equations 3a and 3b.

$$\vec{dH} = \frac{d}{dt} \cdot q_1 \cdot \frac{\vec{ds} \times (\vec{s} - \vec{r})}{4\pi \cdot |\vec{s} - \vec{r}|^3} \qquad (\text{Equ. 3a})$$

$$\vec{H}_{total} = \int_{path} \vec{dH} \qquad (\text{Equ. 3b})$$

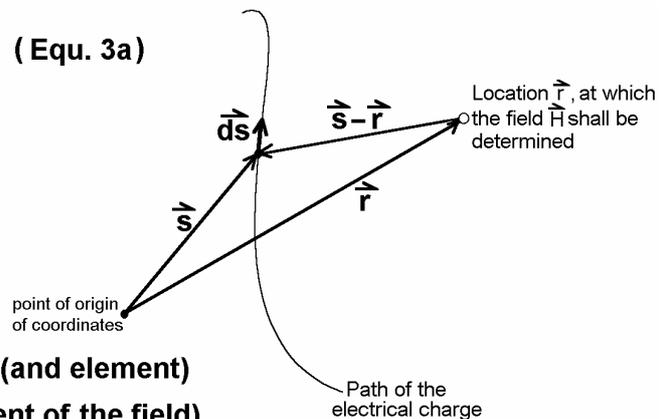

**With:**
$q_1$ = electrical charge
$dt$ = time (infinitesimally small)
$\vec{s}, \vec{ds}$ = path of the charge's movement (and element)
$\vec{dH}$ = magnetic field (infinitesimal element of the field)
$\vec{r}$ = place at which the field has to be determined

**Figure 1:**
**Illustration for the calculation of the magnetic field according to Biot-Savart**

Equation 3a describes the infinitesimal field $d\vec{H}$ of an infinitesimal conductor element $d\vec{s}$ or an infinitesimal part of the path on which the charge $q_1$ is moving. Integration along the path of the movement of the charge (this is the course of the conductor) delivers the total field which emanates from the charge. We see it in equation 3b.

But aim of our interest is not an abstract field but a measurable magnetic force between the moving charges. This force is known under the name Lorentz- force and is calculated according to equation 4a, where $\vec{v}$ is the velocity of the charge $q_2$ relatively to the field $\vec{H}$.

$$\vec{F}_{em} = \mu_o q_2 \cdot (\vec{v} \times \vec{H}) \quad \text{(equ. 4a)}$$

Similar like $\varepsilon_0$ in the static case, also in the dynamic case a factor of proportionality is necessary to connect the sizes of q, $\vec{v}$ and $\vec{H}$) properly with the action, the force $\vec{F}$. In the electrodynamic case, this factor is

$$\mu_0 = 4\pi \cdot 10^{-7} \tfrac{Vs}{Am} \quad \textbf{(equation 4b)}$$

Usually the abbreviation $\quad \vec{B} = \mu_0 \vec{H} \quad$ **(equation 4c)**

is applied, calling $\vec{B}$ magnetic vacuum permeability.
With this abbreviation, the electromagnetic Lorentz- force comes to its familiar expression:

$$\vec{F}_{em} = q \cdot (\vec{v} x \vec{B}) \quad \text{(equation 4d)}$$

In complete analogy with this electromagnetic force, we can now regard the gravimagnetic force. Let us choose the symbols given in table 2 for the physical sizes.

| Electric sizes | Gravitational analogon |
|---|---|
| $\vec{E}$ = electrostatic field | $\vec{G}$ = gravitostatic field |
| $\vec{H}$ = electromagnetic field | $\vec{K}$ = gravimagnetic field |
| $\vec{B}$ = electromagnetic vacuum permeability | $\vec{C}$ = gravimagnetic vacuum permeability |
| I = electric current | J = mass current |
| $\mu_0$ = constant of elecromagnetic vacuum permeability | $\beta$ = constant of gravimagnetic vacuum permeability |
| **Table 2: Choice of the gravimagnetic sizes in analogy to the electrical sizes** ||

To determine the sizes $\vec{K}$, $\vec{C}$, J and $\beta$ we refer to the equations of electrodynamics:

The gravimagnetic field $\vec{K}$, emanating from a moving mass "m" at the position $\vec{r}$ results from equation 5 in analogy to the equations 3a and 3b.

$$\vec{K}(\vec{r}) = \int_{\substack{\text{curve of}\\\text{movement}}} \vec{dK} = \ldots = \int_{\substack{\text{curve of}\\\text{movement}}} m \cdot \frac{\vec{dv} \times (\vec{s}-\vec{r})}{4\pi \cdot |\vec{s}-\vec{r}|^3} \quad \text{(equ. 5)}$$

Equation 6 for the gravimagnetic vacuum permeability $\vec{C}$ corresponds to equation 4c:

$$\vec{C} = \beta \cdot \vec{K} \quad \text{(equ. 6)}$$

Now the force acting on a moving mass $m_2$ in the field of the gravimagnetic vacuum permeability $\vec{C}$ (caused by the mass no.1) is obvious (in analogy to the Lorentz- force):

$$\vec{F}_{em} = q_2 \cdot (\vec{v} \times \vec{B}) \quad \text{(electromagnetic force)}$$
$$\vec{F}_{gm} = m_2 \cdot (\vec{v} \times \vec{C}) \quad \text{(gravimagnetic force)} \quad \text{(equations 7 a\&b)}$$

The only thing still to be done is the determination of the factor of proportionality $\beta$ in equ.6. Our analogy is also sufficient to do this, as displayed in table 3. The only assumption we need is, that the speed of propagation is the same for electromagnetic waves and for gravitational waves – namely the speed of light "c".

As known, this speed of propagation is equal to the product of the static and the magnetic factors of proportionality, as given in equation 7c.

$$\mu_0 \cdot \varepsilon_0 = \frac{1}{c^2} \qquad \text{(equation 7c)}$$

|  | electric case | case of gravitation |
|---|---|---|
| factor of proportionality for calculation of static forces | $\frac{1}{4\pi\varepsilon_o} = 8.893 \cdot 10^9 \frac{Nm^2}{C^2}$ | $\gamma = 6.67 \cdot 10^{-11} \frac{Nm^2}{kg^2}$ |
| factor of proportionality for calculation of dynamic forces | $\mu_o = 4\pi \cdot 10^{-7} \frac{Ns^2}{C^2}$ | $\beta = ?$ |
| relation, conecting static and dynamic factors of proportionality with each other. | $\mu_o \cdot \varepsilon_o = \frac{1}{c^2} \Rightarrow \frac{\mu_o}{\frac{1}{4\pi\varepsilon_o}} = \frac{4\pi}{c^2}$ (see equation 7c) where c = speed of light $c = 2.998 \cdot 10^8 \frac{m}{s}$ | transferred into the image of analogy: $\frac{\beta}{\gamma} = \frac{4\pi}{c^2} \Rightarrow \beta = \frac{4\pi}{c^2} \cdot \gamma$ $\Rightarrow \beta = 9.3255 \cdot 10^{-27} \frac{Ns^2}{kg^2}$ |

table 3.:
**Analogy for the determination of the gravimagnetic vacuum permeability β**

With regard to the very tiny size of the factor of proportionality $\beta$ ($= 9.3255 \cdot 10^{-27} \frac{Ns^2}{kg^2}$), we anticipate, why the gravimagnetic forces are so small. This can be demonstrated with an example as following.

## Part 3.: The example of a gyrator in the gravimagnetic field of the earth

Electromagnetic forces can for instance be measured between conductor loops with electrical currents. Analogously, gravimagnetic forces might be detected between rotating gyroscopes – if the measurement is accurate enough. (This is exactly, what the Gravity- Probe- B Experiment is trying.)

In order to maximize the measurable effect, it is sensible to maximize the mass of this gyroscope, which produces the gravimagnetic field. The largest available gyroscope for mankind is the earth with its rotation (mass $m = 5.985 \cdot 10^{24} kg$, duration of one period of rotation T=86164 sec. relatively to the stars). Consequently we decide to use this earth- gyroscope as source of the gravimagnetic field for our example and we devide our example into the following two steps:

- Determination of the gravimagnetic field of the earth
- Prediction of the reaction of a test- gyroscope onto the gravimagnetic field of the earth

## (3.a) Determination of the gravimagnetic field of the earth

Therefore we have to solve a threedimensional integral with vectorial integrand. The complicity of the integrand makes it necessary to solve the integral numerically by iteration. Thus we have to subdivide the earth into finite elements (numbered with the index "i") and then follow the equations 5 and 8.

$$d\vec{K}_i = dm_i \cdot \frac{\vec{v}_i \times (\vec{s}_i - \vec{r})}{4\pi \cdot |\vec{s}_i - \vec{r}|^3}$$

(equ. 8)

, where i = number of each finite element
$\vec{v}_i$ = velocity of the element no. i
$\vec{s}_i$ = position-vector of the element no. i
$\vec{r}$ = position at which the field $d\vec{K}_i$ is determined

For each element (mass element $dm_i$) the field $d\vec{K}_i$ at the position $\vec{r}$ is determined from equation 8 to be summed up according to equation 5 over the threedimensional volume of the earth. Please see the following four remarks in connection with this work.

(i.) The density of the earth is taken from the so called "Preliminary Reference Earth Model" from geology and geophysics [BUL 85]. The value of the density as a function of the distance from the center of the earth is plotted in figure 2.

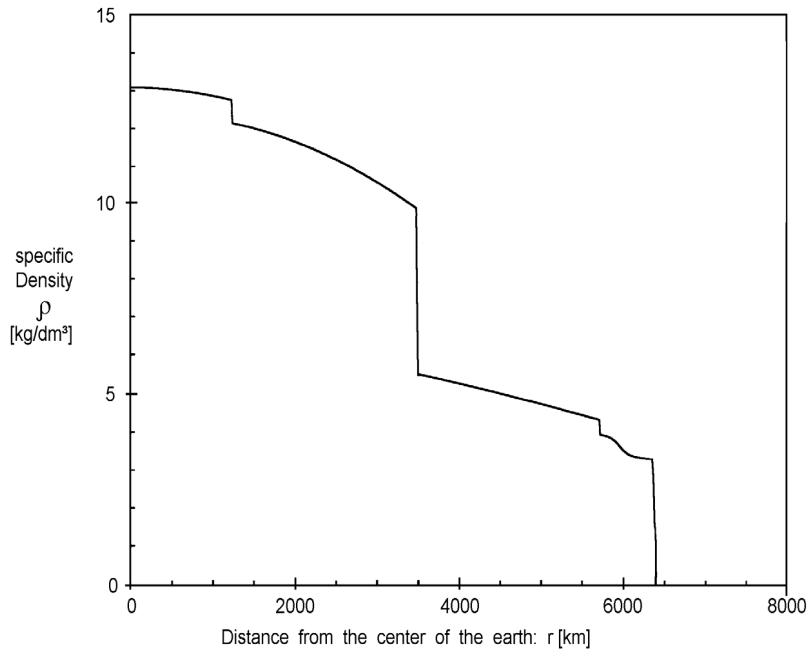

Fig.2: Density of the earth as a function of the distance from the center of the sphere.
The values are taken from the "Preliminary Reference Earth Model" (PREM).

(ii.) From the calculation can be learned, that the gravimagnetic fieldstrength does not depend on the geographical longitude of the position at which the field is calculated. This could be expected because of the rotational symmetry of the earth's movement around the axis connecting the northpole and the southpole. By the way, in this article the degree of latitude is used according to the nomenclature of spherical coordinates in mathematics, which differs from the nomenclature of geograpghy. (Here we have the northpole at θ=0°, the equator at θ=90° and the southpole at θ=180°, see also figure 3). Please keep this in mind when later regarding figure 5.

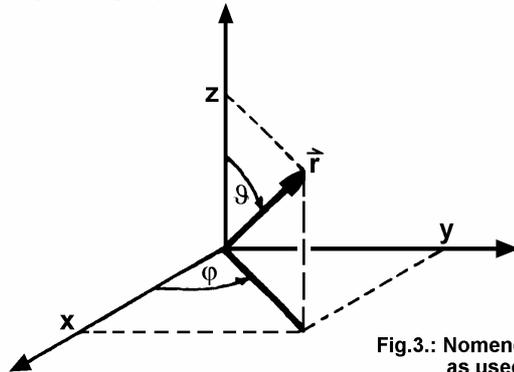

Fig.3.: Nomenclature of spherical coordinates as used in mathematics as well as here.

(iii.) For our calculation it is sufficient to determine the $\vec{K}$ – vector for one arbitrary moment of time, because this vector is fixed rigidly to the sphere of the earth. The total gravimagnetic field of the earth ($\vec{K} = \vec{K}_{ges}$) will be presented in figure 5 according to cartesian coordinates as printed in figure (see also equation 9).

$$\overrightarrow{K_{ges.}} = \sum_i \overrightarrow{dK}_i = \begin{pmatrix} K_x \\ K_y \\ K_z \end{pmatrix} \qquad (Glg.9)$$

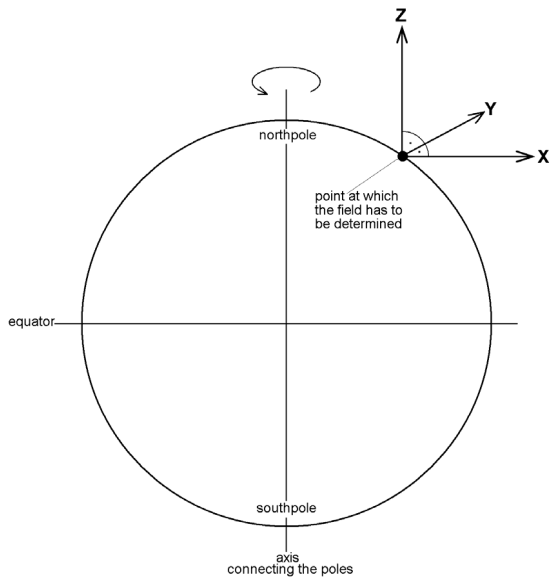

Fig.4.: Orientation of the coordinate-system in which the gravimagnetic field strength is calculated

(iv.) The final presentation of the result (the gravimagnetische field of the earth $\vec{K}$) is shown in figure 5, where we can see the three cartesian components of the field strength and its absolute value as a function of the angle Θ in spherical coordinates.

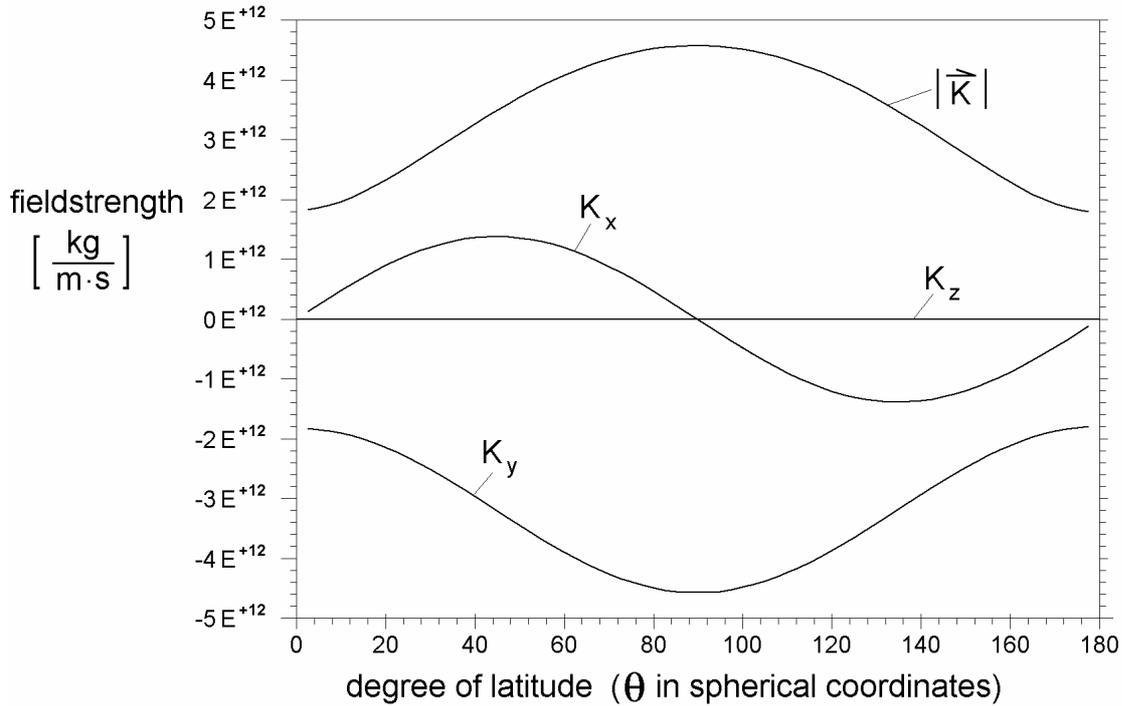

**Fig.5.: Gravimagnetic field strength of the earth (vector $\vec{K}$) as a function of the position in the surface of the earth (by its degree of latitude)**
**The plot gives the cartesian components of $\vec{K}$ and its absolute value.**

This is regarded as gravimagnetic field of the earth. The corresponding gravimagnetic vacuum permeability follows simply by multiplication with the constant $\beta$ according to equation 6.

$$\vec{C} = \beta \cdot \vec{K} = \vec{K} \cdot 9.3255 \cdot 10^{-27} \frac{Ns^2}{kg^2} \qquad \text{(equation 10)}$$

This result brings us into the condition to calculate the example of a gyroscope within the gravimagnetic field of the earth.

### (3.b) The example of a test- gyroscope in the gravimagnetic field of the earth

In the same way as the electromagnetic field of the earth can be verified with a compass, the gravimagnetic field should be verifiable with a gyroscope – presuming the accuracy of the measurement is high enough.
Same as the magnet in a compass can be arbitrarily chosen, the dimensions of the gyroscope can be arbitrarily chosen in a way to optimize the technical conditions for a measurement. Here we want to chose a gyroscope with a shape optimized for the easiness of comprehensibility and calculation - this is cylindersymmetric flywheel. We chose dimension with easy figures:
    Radius of the flywheel = 1 m
    Mass of the flywheel = 1000 kg (one metric ton)
    Time for one rotation = 1 millisec. (per each period)

For our exercise, this is optimum geometry, because it is easy to follow. For the sake of easiness, also the position and the orientation of the flywheel shall be especially chosen. The easiest way is a flywheel located directly on the equator, because the gravimagnetic field there is exactly in negative y-direction. The orientation of the flywheel's axis (its angular momentum) is most easy if parallel to the vector $\vec{r}$ (this is parallel to the x-axis). All these conditions of maximum easiness lead us to an assembly as shown in figure 6.

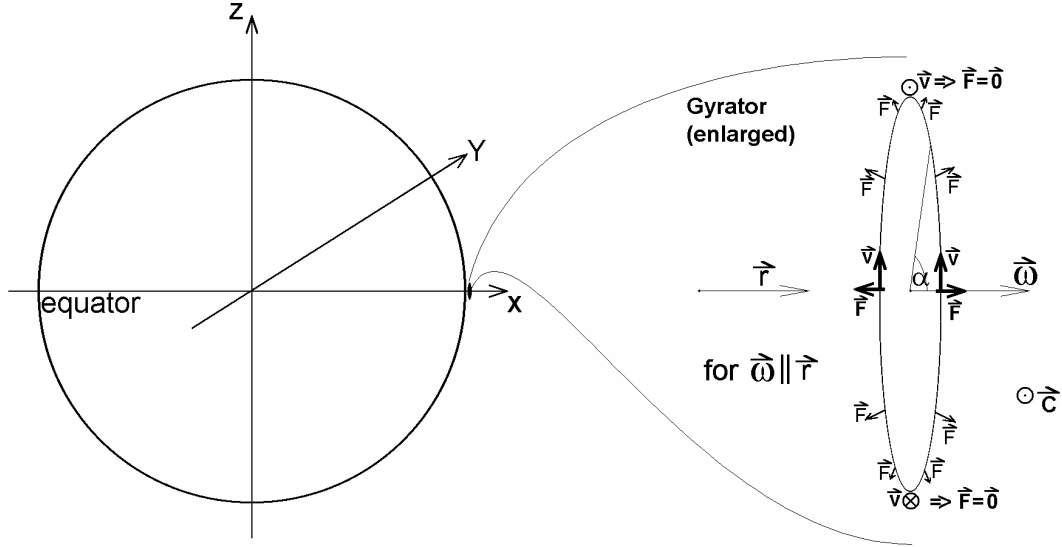

**Fig.6: Gravimagnetic forces, acting on a gyrator- ring, which rotates within the yz-plane. (The vector of angular momentum is orientated parallel to the x-axis)**

The gravimagnetic field $\vec{K}$ and the gravimagnetic vacuum permeability $\vec{C}$ of the earth at this point (equator) are exactly in negative y-direction and have the fieldstrengths:

$$K_y = -4.6 \cdot 10^{+12} \tfrac{kg}{m \cdot sec.} \qquad \textbf{(equation 11a)}$$

and $\qquad C_y = -42.9 \cdot 10^{-15} \tfrac{N \cdot sec.}{kg \cdot m} \qquad \textbf{(equation 11b)}$

For all parts of the gyroscope, the gravimagnetic forces are orientated perpendicular to $\vec{C}$ (see equations 7a and 7b) in analogy with the Lorentz- force. This means, all forces lie within the xz-plane. Because of this, all forces can be summed up rather easy to find out into which type of precession they bring the flywheel. This consideration can be done as following:

Because of the cross product of the vectors $\vec{v}$ and $\vec{C}$ (in Lorentz-force), all force-elements on the flywheel are within the xz-plane and depend on the angle between $\vec{v}$ and $\vec{C}$. At the outermost points of the flywheel ($\alpha=90°$) they reach zero. This means:

$$|\vec{F}| = m \cdot |\vec{v} x \vec{C}| = m \cdot |\vec{v}| \cdot |\vec{C}| \sin(\alpha) = 5.39 \cdot 10^{-10} N \cdot \sin(\alpha) \qquad \text{(equ. 12)}$$

this is per each kilogramm moving mass of the gyrator

Furthermore the effect of this force (this is the tilting moment onto the gyroscopes axis) decreases with the distance of the mass element from the axis. This is simply an effect of the length of the mechanical lever with regard to each element of the flywheel. This length of the lever is the flywheel's radius multipied with sin(α). From there we recieve again a factor of sin(α), which brings us to equation 13.

$$\left|\vec{F}_{eff}\right| = \left|\vec{F}\right| \cdot \sin(\alpha) = m \cdot \left|\vec{v}\right| \cdot \left|\vec{C}\right| \cdot \sin^2(\alpha) \qquad \text{(equation 13)}$$

Putting the value of equation 12 into equation 13, we recieve equation 14:
$$\left|\vec{F}_{eff}\right| = 5.39 \cdot 10^{-10} N \cdot \sin^2(\alpha) \qquad \text{(equation 14)}$$
again per each kilogramm moving mass of the gyroscope

Integration of $\left|\vec{F}_{eff}\right|$ over the circumference of the gyroscope allows us to find the total force acting on the whole gyroscope with a mass of 1000 kg according to equation 15.
$$\left|\vec{F}_{ges}\right| = 2.70 \cdot 10^{-6} N \qquad \text{(equation 15)}$$

This moment of tilt acts on the axis of the flywheel and causes a movement of precession of this flywheel- axis around the direction of $\vec{r}$. As usual we can calculate the frequency of precession as following:

$$\omega_P = \frac{\text{moment of tilt} \quad M}{\text{angular momentum} \quad L} = \frac{2.70 \cdot 10^{-7} kg \cdot \frac{m}{s^2} \cdot m}{2\pi \cdot 10^6 kg \cdot \frac{m}{s^2}} = 4.29 \cdot 10^{-14} \sec^{-1} \qquad \text{(equation 16)}$$

This is the calculated frequency of the Thirring- Lense- rotation in our example.

From these figures we now understand how difficult it is to measure such small forces and such low frequency of precession of such a large gyroscope. The angular frequence of $4.29 \cdot 10^{-14} \frac{rad}{\sec.}$ can be expressed in milli-angular-seconds per year (=marcsec/yr) to get more handy figures: 278 marcsec/yr.

In a real experiment, precondition for the measurement is of course a special bearing of the gyroscope in a way, that earth brings no (or as little as possible) other forces but gravimagnetic forces onto the gyroscope. For instance it is necessary to avoid Coriolis- forces that might overcome gravimagnetic forces easily by several orders of magnitude. In this aspect, it is useful to perform the measurement within a satellite, but unfortunately the frequency of the Thirring- Lense- Rotation ($\omega_P$) decreases with the distance from the center of the earth by a third potential law. This means that the height of the flight of the satellite even decreases the frequency of the precession remarkably. Furthermore the path of the real satellite in the Gravity- Probe- B- experiment is not along the equator, but it is a polar trajectory, going along the north- and south- pole, which additionally decreases the frequency of presession $\omega_P$.

For the Gravity- Probe- B- experiment, the gyroscope is a sphere (by the way it is the most precise sphere manking ever produced), but not a flywheel as in our example. From all these details, we expect that the predicted frequence $\omega_P$ in the Gravity- Probe- B- experiment is smaller than in our simple example. It is only 42 marcsec/yr. The final result of the experiment can be awaited with eagerness.

**Acknowledgement:**
I want to express my thanks to my colleague from the department of mathematics, Prof. Dr. K. Petras, for his extraordinarily friendliness with which he checked my integration according to equation 5 respectively equation 8, which brought me to figure 4.

# Part 4: Reference of Literature: